\documentclass[aps,prl,amsmath,amssymb,english,twocolumn,reprint,floatfix]{revtex4-2}

\usepackage[english]{babel}
\usepackage{graphicx}
\usepackage{verbatim}

\usepackage{epsfig}
\usepackage{epstopdf}
\usepackage{amsfonts}
\usepackage{amsthm}
\usepackage{amsmath}
\usepackage{amssymb}
\usepackage{color}
\usepackage[usenames,dvipsnames,svgnames,table]{xcolor}
\usepackage{hyperref} 
\hypersetup{pdfpagemode=FullScreen,colorlinks=true,linkcolor=NavyBlue,citecolor=BrickRed}
\usepackage{makeidx}
\usepackage{pifont}

\usepackage{xr}
\usepackage{color}
\usepackage{grffile}

\usepackage[ruled]{algorithm2e}
\usepackage[noend]{algpseudocode}
\usepackage{multirow}
\usepackage{type1cm} 
\usepackage{dcolumn}
\usepackage{bm}

\begin{document}

\title{Non-Fermi Liquid and Hund Correlation in La$_4$Ni$_3$O$_{10}$ under High Pressure}

\author{Jing-Xuan Wang}\affiliation{Department of Physics and Beijing Key Laboratory of Opto-electronic Functional Materials $\&$ Micro-nano Devices, Renmin University of China, Beijing 100872, China}\affiliation{Key Laboratory of Quantum State Construction and Manipulation (Ministry of Education), Renmin University of China, Beijing 100872, China}
\author{Zhenfeng Ouyang}\affiliation{Department of Physics and Beijing Key Laboratory of Opto-electronic Functional Materials $\&$ Micro-nano Devices, Renmin University of China, Beijing 100872, China}\affiliation{Key Laboratory of Quantum State Construction and Manipulation (Ministry of Education), Renmin University of China, Beijing 100872, China}
\author{Rong-Qiang He}\email{rqhe@ruc.edu.cn}\affiliation{Department of Physics and Beijing Key Laboratory of Opto-electronic Functional Materials $\&$ Micro-nano Devices, Renmin University of China, Beijing 100872, China}\affiliation{Key Laboratory of Quantum State Construction and Manipulation (Ministry of Education), Renmin University of China, Beijing 100872, China}
\author{Zhong-Yi Lu}\email{zlu@ruc.edu.cn}\affiliation{Department of Physics and Beijing Key Laboratory of Opto-electronic Functional Materials $\&$ Micro-nano Devices, Renmin University of China, Beijing 100872, China}\affiliation{Key Laboratory of Quantum State Construction and Manipulation (Ministry of Education), Renmin University of China, Beijing 100872, China}

\date{\today}

\begin{abstract}
High temperature superconductivity was recently found in the bilayer nickelate $\rm{La}_3 \rm{Ni}_2 \rm{O}_7$ (La327), followed by the discovery of superconductivity in the trilayer $\rm{La}_4 \rm{Ni}_3 \rm{O}_{10}$ (La4310), under high pressure. Through studying the electronic correlation of La4310 with DFT+DMFT, and further comparing it with that of La327, we find that the $e_g$ orbitals of the outer-layer Ni cations in La4310 have a similar (but slightly weaker) electronic correlation to those in La327, in which the electrons behave as a non-Fermi liquid with Hund correlation and linear-in-temperature scattering rate. Our results suggest that the experimentally observed ``strange metal'' behavior may be explained by the Hund spin correlation featuring high spin states and spin-orbital separation. In contrast, the electrons in the inner-layer Ni cations in La4310 behave as a Fermi liquid. The weaker electronic correlation in La4310 is attributed to more hole-doping, which may explain its lower superconducting transition temperature.
\end{abstract}


\maketitle

Recently, the experimental discovery of La327 high-temperature superconductor with a maximum transition temperature ${(T_c)}$ of 80 K under high pressure has attracted considerable attention \cite{sun2023nature}. Due to the limitation of high pressure conditions, only a limited number of experimental works have been studied in the synthesis of crystals and the measurement of electrical transport for La327 \cite{zhou2023evidence,wang2023pressure,wang2023structure}. While, quite a number of theoretical studies in La327 have captured various aspects, such as interlayer antiferromagnetic correlation between Ni 3$d_{z^2}$ and O 2$p_{z}$ orbitals \cite{luo2023prl,ryee2023critical,shi2023prb}, charge density wave order \cite{chen2023evidence,liu2023prl}, bonding state metallization of Ni 3$d_{z^2}$ orbitals \cite{liu2023electronic}, Hund correlation \cite{tian2023correlation,ouyang2023hund,cao2023flat,kakoi2023pair}, superconducting pairing symmetry \cite{gu2023effective,liu2023role,zhang2023prb}, and ground state magnetism \cite{zhang2024prb,zhang2023structural}.

The nickelate Ruddlesden-Popper (RP) phases La$_{n+1}$Ni$_{n}$O$_{3n+1}$ are of perovskite-like structure with ${n}$-layer $\rm{NiO_6}$ octahedrons. The NiO$_2$ layers in RP phase is structurally similar to the $\rm{CuO_2}$ layer in high-${T_c}$ cuprate superconductors \cite{cu1989pc,cu1991ansaldo,cukim2013}. The bilayer La327 $(n=2)$ shares a similar structure to the trilayer La4310 $(n=3)$, where the nominal valence states of Ni atoms are +2.5 and +2.67, respectively, so they may have similar physical properties, especially unconventional superconductivity \cite{leonov2024electronic,sakakibara2023theoretical}. In addition, La4310 undergoes a structural transition from the monoclinic ${P2{1}/a}$ to the tetragonal ${I4/mmm}$ phase in the pressure range of 12.6 to 13.4 GPa \cite{li2023structural}. 

In the past years, studies on La4310 primarily have been focused on the metal-to-metal phase transition at approximately 130 to 160 K \cite{zhang2017np,zhang2020nc,zhang2020prm,huang2020prb,ning2024jcg}. More recently, several experiments have attempted to find superconductivity in La4310 \cite{li2023cpl,li2023structural,zhang2023superconductivity}. In the experiment conducted by Zhu et al. \cite{zhu2023signatures}, it is noteworthy that zero resistance was observed and superconductivity was demonstrated with a maximum ${T_c}$ of approximately 30 K in La4310. Meanwhile, ``strange metal” behavior in La4310 was found with a linear temperature-dependent resistance extending up to 300 K, which may be linked to the enhanced high spin fluctuation and strong correlation. The correlations in the bilayer La327 and the trilayer La4310 may have profound impact on their physical properties, such as electronic structures, superconducting pairing mechanism, conductivity, and high-temperature superconductivity. However, an understanding of the correlation effect in La4310 is still lacking. Thus, it is of importance to study the physical properties by considering correlations in La4310. Comparing the correlations in La327 and La4310 as well as two non-equivalent NiO$_2$ layers in La4310 will not only have a potential to elucidate why the ${T_c}$ of La4310 is lower than that of La327, but also provide a perspective on the underlying superconducting mechanism.

In this Letter, we investigate the electronic correlation in La4310 and further compare it to that in La327 using density functional theory based calculations combined with dynamical mean-field theory calculations (DFT+DMFT). We find that the correlations are layer-dependent as well as orbital-dependent. The underlying reason is the different occupation numbers and hence different valences for the Ni cations, which implies different hole dopings. The inner NiO$_2$ layer is more hole doped and weakly correlated, where the Ni cations have a higher valence. Similar to the NiO$_2$ layer in La327, the outer NiO$_2$ layers in La4310 are more correlated and behave as a non-Fermi liquid with a characteristic $T$-linear scattering rate for the electrons, given rise to by the Hund spin correlation indicated by high-spin states and spin-orbital separation. Compared to La327, the slightly weaker correlation in the outer NiO$_2$ layers and the Fermi-liquid middle NiO$_2$ layer in La4310 may explain its lower superconducting $T_c$.


\begin{figure*}[htb]
  \centering
  \includegraphics[width=17.2cm]{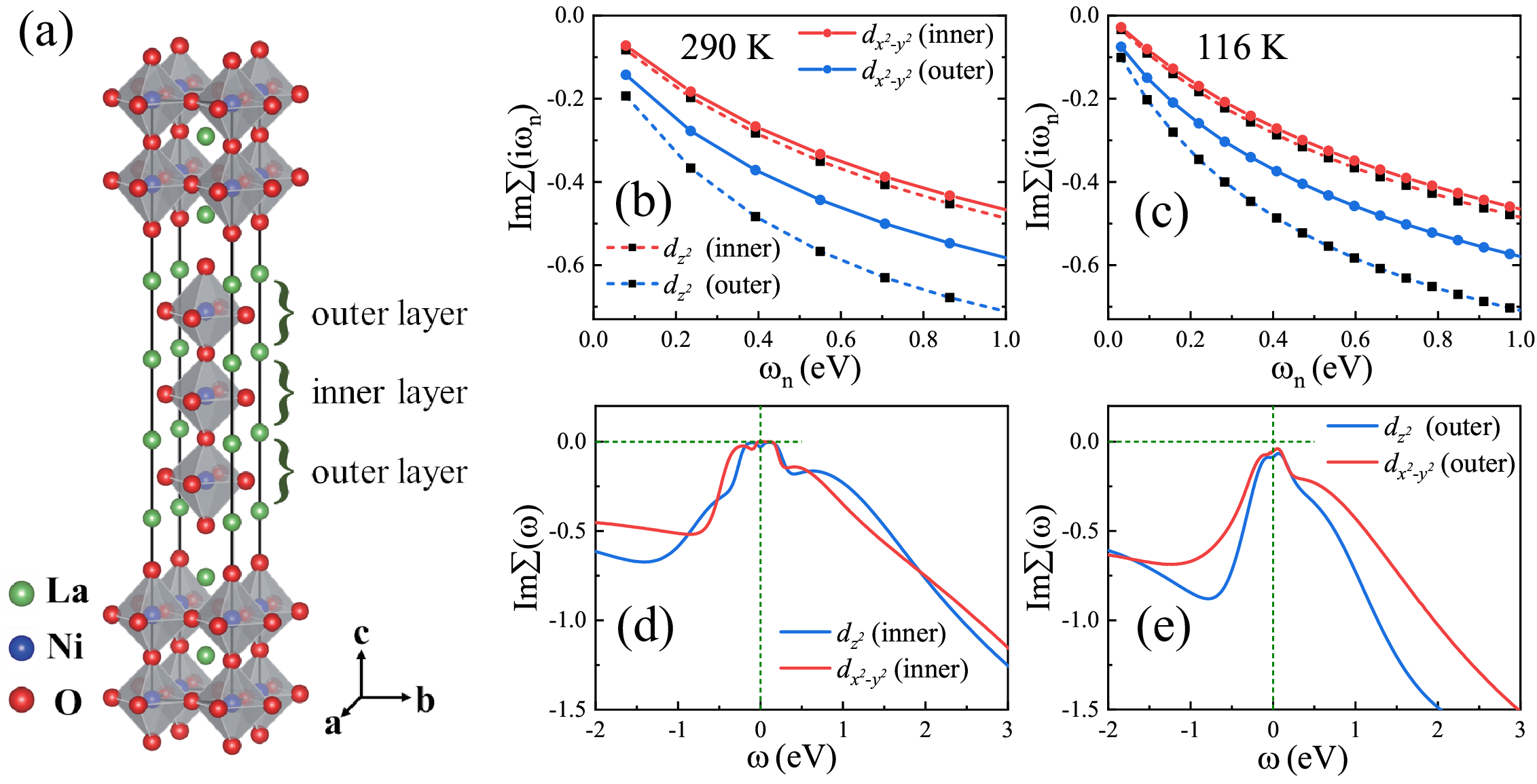}
  \caption{(a) The crystal structure of $\rm{La}_4 \rm{Ni}_3 \rm{O}_{10}$. (b), (c) The imaginary parts of the self-energy functions at the Matsubara axis $\text{Im}\Sigma(i\omega_n)$ at 290 and 116 K, respectively. (d), (e) The real frequency $\text{Im}\Sigma(\omega)$ for the inner- and outer-layer Ni atoms at 290 K.}
  \label{fig:structure}
\end{figure*}

{\it Method.} We carried out charge self-consistent DFT+DMFT calculations using the code EDMFT, based on WIEN2K \cite{Blaha2020jcp}. Only the two ${e_g}$ ($d_{z^2}$ and $d_{x^2-y^2}$) orbitals of Ni were considered to be correlated, while the three ${t_{2g}}$ orbitals were filled and hence not included in the DMFT calculations. Since the Ni atoms have two nonequivalent Wyckoff positions in the La4310 structure, we considered the correlations of the inner- and outer-layer Ni atoms as two impurity problems in DMFT. ${U}$ = 5.0 $\rm{eV}$ and ${J_H}$ = 1.0 $\rm{eV}$ were chosen in the calculations, which are the typical values used previously for La327 \cite{cao2023flat, ouyang2023hund}. For further details of the method, see \cite{sm}.


{\it Results.} We studied the newly reported RP phase of the nickelate family, focusing on the trilayer La4310 under high pressure of 44.3 $\rm{GPa}$, whose structure belongs to the space group ${I4/mmm}$ (No.139) with lattice parameters $a = 3.661\, \text{\AA}$, $c = 26.278\, \text{\AA}$ \cite{li2023structural}. Fig.~\ref{fig:structure}(a) displays the crystal structure and ${\rm NiO_6}$ octahedrons. There are three layers of Ni atoms in the middle of the two rock-salt layers (double La and O layers). We define the upper and lower layers as the outer Ni layers and the middle layer as the inner Ni layer.

\begin{figure}[htb]
  \centering
  \includegraphics[width=8.6cm]{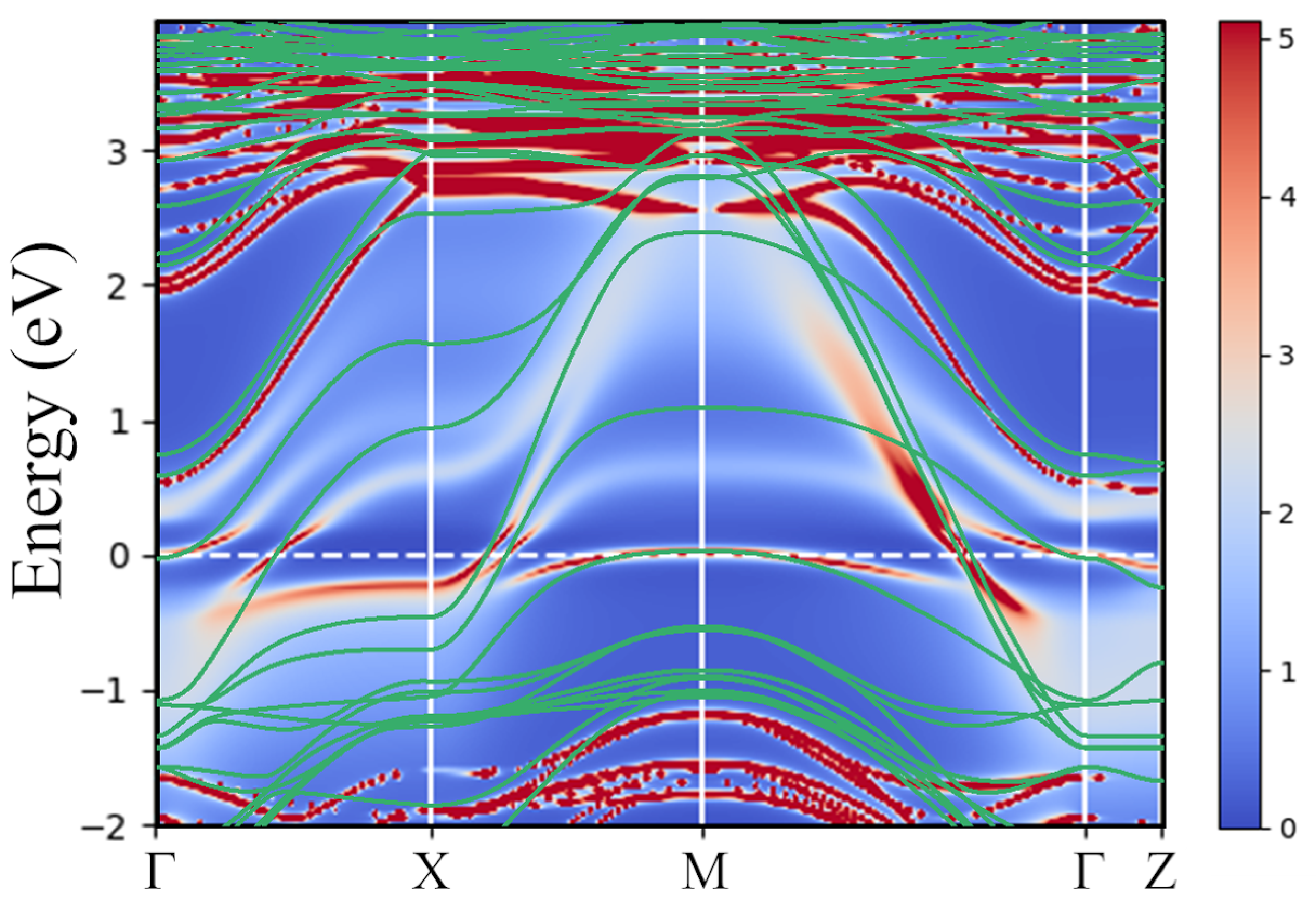}
  \caption{Momentum-resolved spectral functions $A({\mathbf{k}}, w)$ of La4310 by DFT+DMFT calculated at 290 K. The DFT band structure is plotted with green lines for reference. The Fermi level is set to zero energy as marked by the white dashed line. The energy bands of the Ni ${e_g}$ orbitals are mainly in the energy range of $-$1 to 3 eV, which are somewhat blurry and narrowed by the correlation-induced renormalization. The flat bands are clearly visible at the Fermi level (near the M point), which are dominantly from the Ni $3d_{z^2}$ orbitals.}
  \label{fig:band}
\end{figure}

{\it Orbital-dependent and layer-dependent electronic correlation.} We calculated the self-energy of the correlated two $e_g$ (Ni 3$d_{z^2}$ and 3$d_{x^2-y^2}$) orbitals of the inner and outer Ni layers. The imaginary parts of the Matsubara-frequency self-energy $\text{Im}\Sigma(i\omega_n)$ are shown in Fig.~\ref{fig:structure}(b) and (c) at 290 and 116 K, respectively. In Fig.~\ref{fig:structure}(d) and (e), we plot the imaginary parts of the real-frequency self-energy $\text{Im}\Sigma(\omega)$ at 290 K. For both the inner- and outer-layer Ni atoms, $\text{Im}\Sigma(i\omega_n)$ magnitudes of the $d_{z^2}$ orbitals are larger than those of the $d_{x^2-y^2}$ orbitals. Moreover, the slopes of $\text{Im}\Sigma(i\omega_n)$ at low frequencies for the $d_{z^2}$ orbitals are larger than those for the $d_{x^2-y^2}$ orbitals. These results show that the $d_{z^2}$ orbitals have stronger correlation than the $d_{x^2-y^2}$ orbitals, which is also revealed by that the energy bands of the $d_{z^2}$ orbitals are flatter and narrower than those of the $d_{x^2-y^2}$ orbitals, as shown from the momentum-resolved spectral functions $A({\mathbf{k}}, w)$ in Fig.~\ref{fig:band}.


On the other hand, for both the outer-layer Ni ${e_g}$ orbitals, the magnitudes of their $\text{Im}\Sigma(i\omega_n)$ are larger than those of the inner-layer, and the corresponding intercepts of their $\text{Im}\Sigma(\omega)$ can be observed around zero frequency, indicating the outer-layer Ni atoms have stronger correlation than the inner-layer ones, as shown in Fig.~\ref{fig:structure}(e). For the inner-layer Ni ${e_g}$ orbitals, their $\text{Im}\Sigma(i\omega_n)$ at low frequencies are of linear behavior and the corresponding $\text{Im}\Sigma(\omega)$ are close to 0 around zero frequency (Fig.~\ref{fig:structure}(d)), which is consistent with a Fermi liquid behavior, especially for the $d_{x^2-y^2}$ orbital. In comparison with La327, the magnitudes of $\text{Im}\Sigma(i\omega_n)$ for the inner- and outer-layer are all smaller than those of La327, which indicates that the Ni atoms in La327 have the strongest correlation. This may explain that La327 has a higher superconducting ${T_c}$ than La4310.

Fig.~\ref{fig:band} shows the momentum-resolved spectral functions $A({\mathbf{k}}, w)$ of the trilayer La4310 at 290 K. To further clarify the effect of temperature, we show temperature-dependence density of states in Fig.~\ref{fig:s1} \cite{sm}. The flat band (near the M point) at the Fermi level is clearly visible, mainly from the ${\rm Ni}$ $3d_{z^2}$ orbitals, and manifests a strong band renormalization. The bandwidth of the $e_g$ orbitals in La4310 is almost the same as that in La327. Compared to La327, the energy bands of La4310 are clearer, showing stronger coherence and weaker correlation in La4310.

\begin{table}[t]
  \begin{center}
  \renewcommand\arraystretch{1.5}
  \caption{Local occupation numbers ${N_d}$ of the $d_{z^2}$ and $d_{x^2-y^2}$ orbitals for the inner-layer, outer-layer, and La327 \cite{ouyang2023hund} at 290 K.}
  \label{tab1}
  \begin{tabular*}{8.6cm}{@{\extracolsep{\fill}} cccccc}
  \hline\hline
    &orbitals & inner & outer& La327\\
  \hline 
  \multirow{3}*{$N_d$} & $d_{z^2}$ & 1.060 &  1.088 &1.139 \\ 
   & $d_{x^2-y^2}$ & 1.017  & 1.060 &  1.056 \\
   & total & 2.077& 2.148& 2.195\\
 
  \hline\hline
 \end{tabular*}
 \end{center}
 \end{table}

The average valences of Ni in La4310 and La327 are +2.67 and +2.5, which give the nominal 3${d^{7.33}}$ and 3${d^{8}}$ configurations for the Ni atoms, respectively \cite{lu2023interlayer, oh2023type,jiang2024high}. The configuration of half-filled ${d_{z^2}}$ orbitals and quarter-filled ${d_{x^2-y^2}}$ orbitals with the ${e_g}$ orbitals occupation of 1.5 for Ni atoms in La327 is suggested by simple electron count \cite{sun2023nature}. However, the occupation numbers of the ${e_g}$ orbitals are more than 2 as shown in Table~\ref{tab1}. There is a notable difference from our calculations. The origin of this enhancement in the local occupancy number is the orbital hybridizations between the Ni-${3d}$ and O-${2p}$ orbitals, which gives rise to charge transfer from $\rm {O^{2-}}$ to $\rm {Ni^{2.67+}}$. In general, orbital hybridizations are almost inevitable, which usually leads to a different occupancy than the nominal one from chemical valence analysis. This behavior can also be found in other nickelates. For example, the DFT+DMFT calculations for ${\rm SrNiO_2}$ show that the calculated occupation of Ni-${3d}$ is 8.479, which is inconsistent with that of the nominal configuration of 3${d^{8}}$ \cite{wang2020prb}. Since the LaO layers are insulating and each La-O atom pair has a valence of $+$1, each O-Ni-O in the NiO$_2$ layers has a valence of $-$1.33 and the NiO$_2$ layers are essentially 0.77 hole doped compared with their half-filling configuration composed of Ni-3$d^8$ and O-2$p^6$, although the occupation number of each ${e_g}$ orbitals of Ni atoms is greater than 1.

The occupation numbers of the ${d_{z^2}}$ orbitals in the above three types of Ni atoms are all more than the ones of the ${d_{x^2-y^2}}$ orbitals, which means that the ${d_{x^2-y^2}}$ orbitals have more hole doping and partially account for the weaker correlation. The occupation numbers of the inner-layer ${e_g}$ orbitals are less than those of the outer-layer, indicating that the inner layer has more hole doping and its Ni cations have a higher valence than +2.67. This is consistent with our analysis above that the inner-layer Ni atom behaves as a Fermi liquid. In comparison with the case of La327, the outer-layer Ni atoms have a similar occupation number for the ${d_{x^2-y^2}}$ orbitals, while 0.051 less for the ${d_{z^2}}$ orbitals. It implies more hole doping in the ${d_{z^2}}$ orbitals of La4310, which would weaken the interlayer antiferromagnetic correlation of the ${d_{z^2}}$ orbitals and further affect that of the ${d_{x^2-y^2}}$ orbitals and hence is adverse to superconductivity \cite{tian2023correlation,luo2023prl,shi2023prb}.

\begin{figure}[tb]
  \centering
  \includegraphics[width=6.0cm]{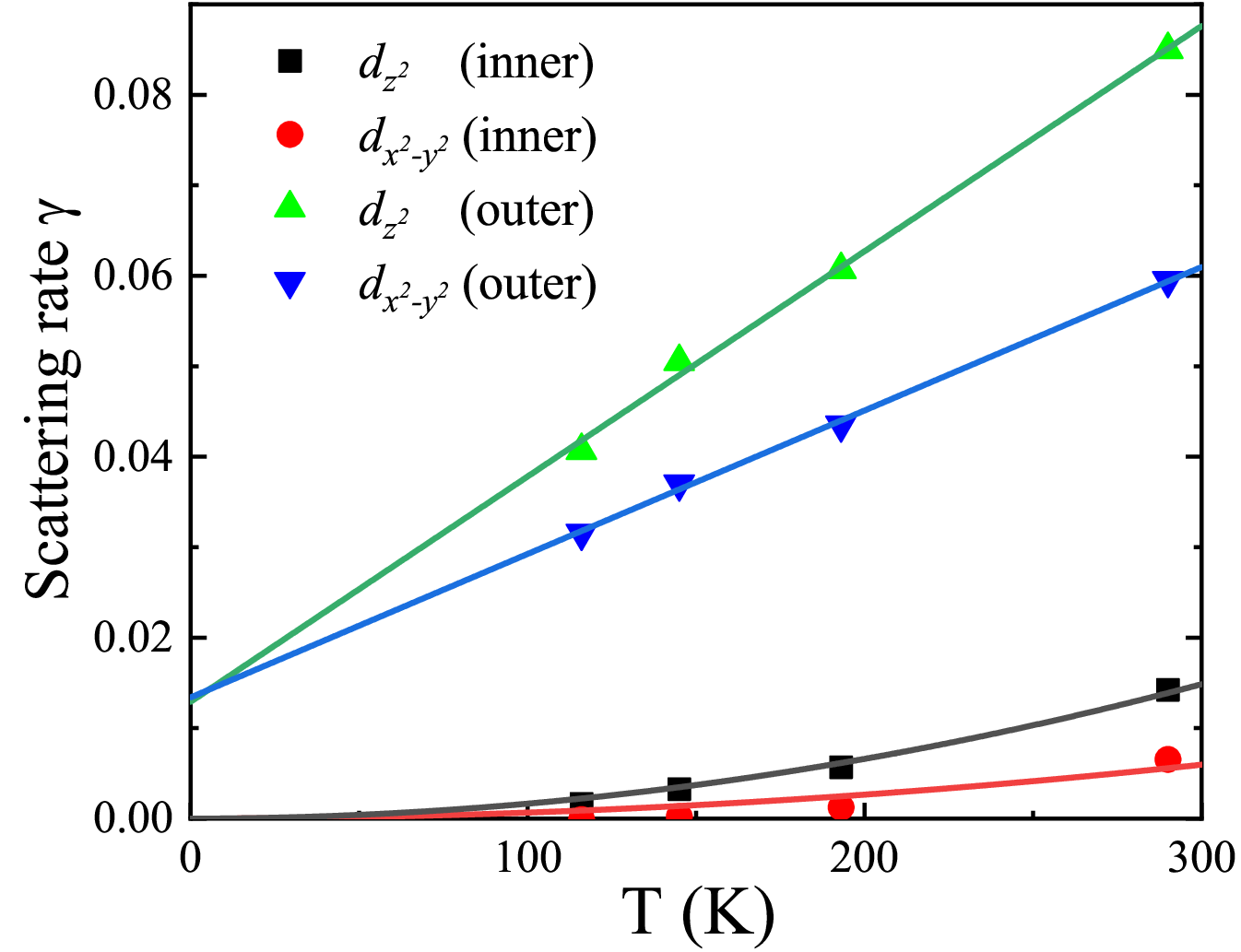}
  \caption{The imaginary part of the self-energy at zero frequency, $\rm{-Im}\Sigma(\omega=0)$, plotted as functions of temperature by DFT+DMFT calculated. The green and blue lines are temperature linear fit for the outer-layer Ni ${e_g}$ orbitals that exhibit ``strange metal'' and non-Feimi liquid behavior at low temperatures. The red and black lines are temperature squared $(\sim{T^2})$ fit for the inner-layer ${e_g}$ orbitals, showing Fermi-liquid behavior.}
  \label{fig:iw}
\end{figure}

{\it ${T}$-linear scattering rate.} Then, we studied the electronic scattering rate ${\gamma}\equiv{\rm -Im\Sigma}{(\omega=0)}$ for the different orbitals. It corresponds to the inverse of the quasiparticle lifetime \cite{varma1989prl,Parcollet1999prb,schroder2000nature}. The imaginary part of the self-energy at zero frequency is obtained from a second-order polynomial extrapolation for the self-energy on the Matsubara frequency, ${\gamma}$ ${\approx}$ $-{\rm Im}$$[1.875\Sigma(i\omega_0)$ $-$ $1.25\Sigma(i\omega_1)$ $+$ $0.375\Sigma(i\omega_2)]$ \cite{wuwei2022}. The scattering rate ${\gamma}$ with temperatures from 116 to 290 K are shown in Fig.~\ref{fig:iw}. It shows $T$-quadratic behaviors for the inner-layer ${e_g}$ orbitals at low temperatures, which is characteristic of a Fermi liquid. The scattering rate of the outer-layer ${e_g}$ orbitals can be well fitted with a linear-in-temperature function, indicating that the lifetime of the quasiparticle is inversely proportional to temperature, exhibiting a non-Fermi liquid and a ``strange metal'' behavior.

{\it Hund correlation.} The ``strange metal'' behavior for La4310 may originate from the Hund correlation effect. Table~\ref{tab2} presents the probabilities of all local spin multiplets for the inner-layer, outer-layer, and La327 Ni atoms at 290 K. The high-spin state of ${N_\Gamma=2}$ and $S^z_\Gamma=1$ has the largest weight, showing that the spins of the ${e_g}$ orbitals tend to align parallel. Apparent high-spin state is a consequence of the Hund coupling and we refer to it as Hund spin correlation. Note that the high-spin state $S^z_\Gamma=1$ for La327 has the largest probability with 33.3${\%}$, followed by the outer layer in La4310 with 31.5${\%}$, and the inner layer with 27.6${\%}$, which indicates that La327 has the strongest Hund correlation and the inner layer in La4310 has the weakest Hund correlation, which is quite consistent with the correlations signified by the self-energies and the hole-doping mechanism.

\begin{table}[t]
  \begin{center}
  \renewcommand\arraystretch{1.5}
  \caption{Weights of the local multiplets for the inner- and outer-layer $e_{g}$ orbitals for La4310 at 290 K. The last line shows the results for La327 calculated with the same method \cite{ouyang2023hund}. ${N_\Gamma}$, ${S^z_\Gamma}$, ${P_\Gamma}$ represent total occupation number, total spin and probability of the local multiplets, respectively.}
  \label{tab2}
  \begin{tabular*}{8.6cm}{@{\extracolsep{\fill}} ccccccc}
  \hline\hline
  ${N_\Gamma}$ & 0 & 1 & 2 & 2 & 3  &4\\
  ${S^z_\Gamma}$ & 0 & 1/2 & 0 & 1 &1/2 & 0 \\
  \hline
  ${P_\Gamma}$ (inner) & 1.0${\%}$	& 18.3${\%}$ &	27.1${\%}$	& 27.6${\%}$ &	24.1${\%}$ &	1.9${\%}$\\
  ${P_\Gamma}$ (outer) & 0.6${\%}$	&	14.7${\%}$	&	24.8${\%}$	&	31.5${\%}$	&	26.4${\%}$	&	2.1${\%}$	\\
  ${P_\Gamma}(\rm La327)$ & 0.4${\%}$	& 12.4${\%}$	& 23.6${\%}$	& 33.3${\%}$	& 28.1${\%}$	& 2.3${\%}$	\\
  \hline\hline
\end{tabular*}
\end{center}
\end{table}

\begin{figure}[htb]
  \centering
  \includegraphics[width=8.6cm]{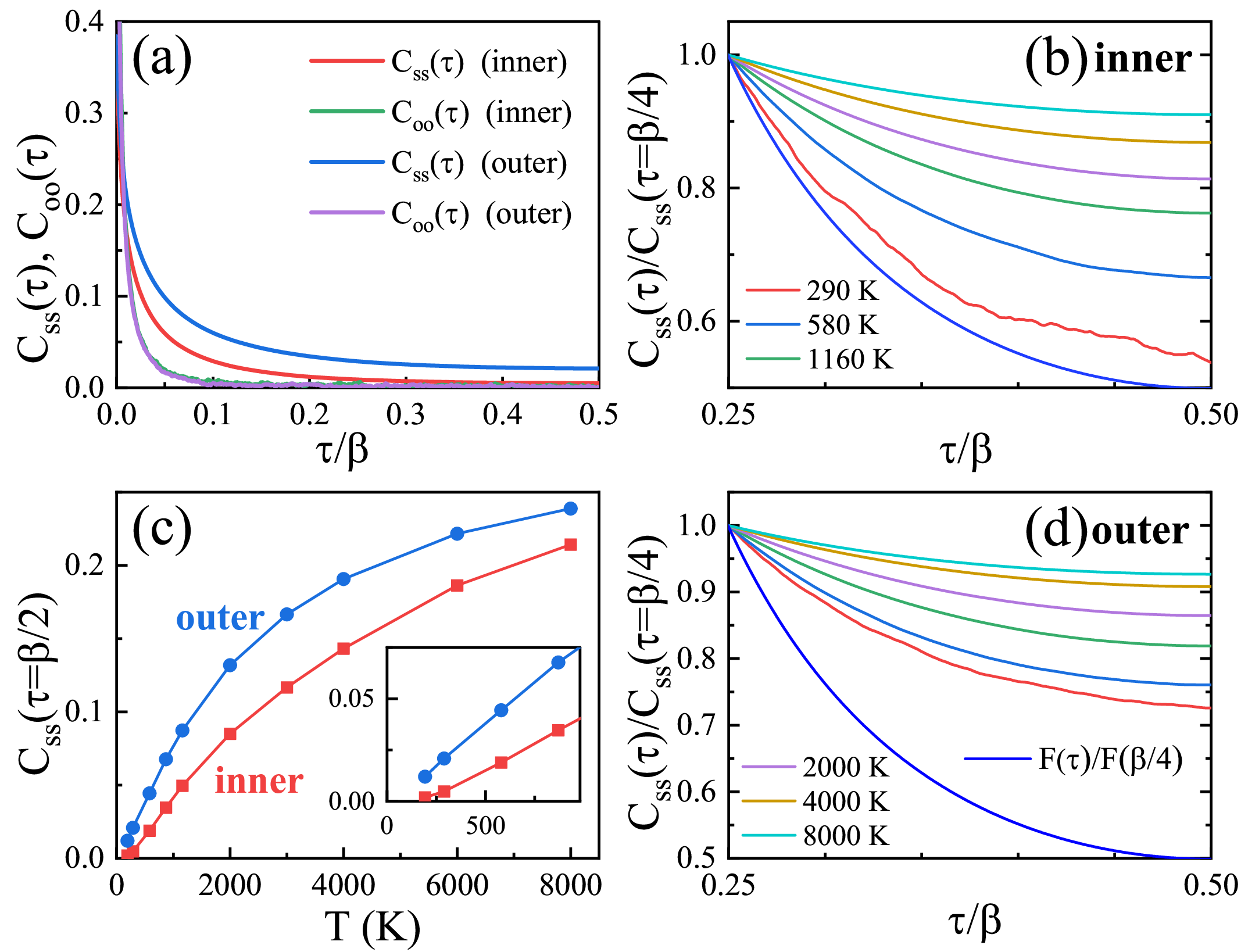}
  \caption{(a) The imaginary-time spin-spin correlation functions $C_{\text{ss}}(\tau)$ and orbital-orbital correlation functions $C_{\text{oo}}(\tau)$ for the ${e_g}$ orbitals at 290 ${\rm K}$. (c) $C_{\text{ss}}(\tau ={\beta}/{2})$ plotted as functions of temperature. The inset in (c) is an enlarged view of the low-temperature range. (b), (d) $C_{\text{ss}}(\tau)/C_{\text{ss}}(\tau=\beta/4)$ and $F(\tau)/F(\beta/4)$ plotted as functions of temperatures.}
  \label{fig:susc2}
\end{figure}

To reveal the physical origin of the ${T}$-linear scattering rate, we further studied the imaginary-time spin-spin correlation functions $C_{\text{ss}}(\tau)=\langle S_z(\tau)S_z(0) \rangle$ and orbital-orbital correlation functions $C_{\text{oo}}(\tau)=\langle O(\tau)O(0) \rangle$, where $S_z$ is the total spin, and ${O}$ is the charge difference $n(d_{z^2})-n(d_{x^2-y^2})$. In Fig.~\ref{fig:susc2}(a), $C_{\text{oo}}(\tau)$ decays substantially faster than $C_{\text{ss}}(\tau)$, showing spin-orbital separation, which is a typical feature for Hund metal. The same feature is also illustrated in the local magnetic susceptibility in Fig.~\ref{fig:s2} \cite{sm}. The spin-orbital separation in Hund systems has been found previously \cite{stadler2015prl,stadler2019ap,yin2012prb,deng2019nature,Werner2008prl}. $C_{\text{ss}}(\tau)$ for the outer-layer decays to a finite residual value at long times, indicating the spins have not been fully screened. Fig.~\ref{fig:susc2}(c) shows $T$-linear behavior for the residual value in a temperature range from 290 to 870 K, which may be related to the ${T}$-linear scattering rate found above. 

The temperature dependence of $C_{\text{ss}}(\tau)/C_{\text{ss}}(\tau=\beta/4)$ and $F(\tau)/F(\beta/4)$ \cite{Werner2008prl}, where $F(\tau)=(T/sin(\pi \tau T))^2$, for the inner- and outer-layer Ni atoms is depicted in Fig.~\ref{fig:susc2}(b) and (d). The local magnetic moment is not completely screened at high temperatures and is gradually screened as the temperature decreases. For the inner layer, $C_{\text{ss}}(\tau)$ behaves like $F(\tau)$ below 290 K, which is also an evidence for Fermi liquid.

Wu et al. \cite{wuwei2022} discovered that the antiferromagnetic fluctuations are the origin of ${T}$-linear scattering rate and strange metallicity in the single-band Hubbard model with cluster DMFT. Here, in La4310 and La327 the orbital degrees of freedom are screened completely \cite{sm,ouyang2023hund}, while the spins are only partially screened. Therefore, we argue that the Hund spin correlation, as we observed, is the origin of the ${T}$-linear scattering rate and ``strange metal”. The Hund correlation and spin alignment in the ${e_g}$ orbitals are important for both correlation and superconductivity, which has been confirmed in many studies \cite{tian2023correlation, cao2023flat, wang2020prb, zhao2021prb}. 

{\it Conclusion.} We have investigated the non-Fermi liquid and Hund correlation in La4310 using DFT+DMFT. Our findings indicate that the inner layer possesses weak Hund correlation and a Fermi liquid behavior, mainly because of more hole doping and a higher valence than +2.67. The outer layers, similar to the $\rm{NiO_2}$ layers in La327, possess strong Hund correlation, which gives rise to a non-Fermi liquid. In addition, the outer layers have a $T$-linear scattering rate, may explain the “strange metal” observed in experiments, whose origin is suggested to be the Hund spin correlation. Compared with La327, the slightly weaker electronic correlation in La4310 may explain its lower superconducting $T_c$.

\begin{acknowledgments}
  This work was supported by National Natural Science Foundation of China (Grant No. 11934020). Computational resources were provided by Physical Laboratory of High Performance Computing in RUC.
\end{acknowledgments}

\bibliography{lno4310hund}

\end{document}


\title{Supplementary Information for: \\Non-Fermi Liquid and Hund Correlation in La$_4$Ni$_3$O$_{10}$ under High Pressure}
\author{Jing-Xuan Wang}\affiliation{Department of Physics and Beijing Key Laboratory of Opto-electronic Functional Materials $\&$ Micro-nano Devices, Renmin University of China, Beijing 100872, China}\affiliation{Key Laboratory of Quantum State Construction and Manipulation (Ministry of Education), Renmin University of China, Beijing 100872, China}
\author{Zhenfeng Ouyang}\affiliation{Department of Physics and Beijing Key Laboratory of Opto-electronic Functional Materials $\&$ Micro-nano Devices, Renmin University of China, Beijing 100872, China}\affiliation{Key Laboratory of Quantum State Construction and Manipulation (Ministry of Education), Renmin University of China, Beijing 100872, China}
\author{Rong-Qiang He}\email{rqhe@ruc.edu.cn}\affiliation{Department of Physics and Beijing Key Laboratory of Opto-electronic Functional Materials $\&$ Micro-nano Devices, Renmin University of China, Beijing 100872, China}\affiliation{Key Laboratory of Quantum State Construction and Manipulation (Ministry of Education), Renmin University of China, Beijing 100872, China}
\author{Zhong-Yi Lu}\email{zlu@ruc.edu.cn}\affiliation{Department of Physics and Beijing Key Laboratory of Opto-electronic Functional Materials $\&$ Micro-nano Devices, Renmin University of China, Beijing 100872, China}\affiliation{Key Laboratory of Quantum State Construction and Manipulation (Ministry of Education), Renmin University of China, Beijing 100872, China}

\date{\today}

\maketitle
\renewcommand{\thepage}{S\arabic{page}}  
\renewcommand{\thesection}{S\arabic{section}}   
\renewcommand{\thetable}{S\arabic{table}}   
\renewcommand{\thefigure}{S\arabic{figure}}
\renewcommand{\theequation}{S\arabic{equation}}


\section{Method}\label{sec:DMFT}

We carried out charge self-consistent DFT+DMFT calculations using the code EDMFT, based on WIEN2K \cite{Blaha2020jcp}. The exchange-correlation energy was described with the generalized gradient approximation (GGA) and the Perdew-Burke-Ernzerhof (PBE) functional \cite{Perdew1996prl}. The muffin tin radii $R_{MT}$ were 2.29, 1.85, and 1.59 bohr for La, Ni, and O, respectively. The maximum modulus for the reciprocal vectors $K_{\max }$ was chosen such that $R_{M T} \times K_{\max }=7.0$. The ${k}$-mesh was used with $7 \times 7 \times 7$ in the first Brillouin zone, which was set to $68 \times 68 \times 68$ when calculating the density of states. A large energy window $\mathcal{-}$10 to 10 $\rm{eV}$ with respect to the Fermi level was chosen construct localized 3${d}$ orbitals of Ni atom. The impurity is solved by the hybridization expansion version of the continuous-time quantum Monte Carlo (CTQMC) with exact double-counting scheme for the self-energy function \cite{Gull2011rmp,Haule2015prl}. We set the Monte Carlo step to $3 \times 10^7$ and we checked the convergence of self-energy and energy after the calculation was completed. Then, the self-energy on the real frequency was obtained by the analytical continuation method of maximum entropy \cite{Gull2011rmp}. We neglected the spin-orbital coupling effects.

\section{Density of states}\label{sec:DOS}
\begin{figure}[htb]
  \centering
  \includegraphics[width=8.6cm]{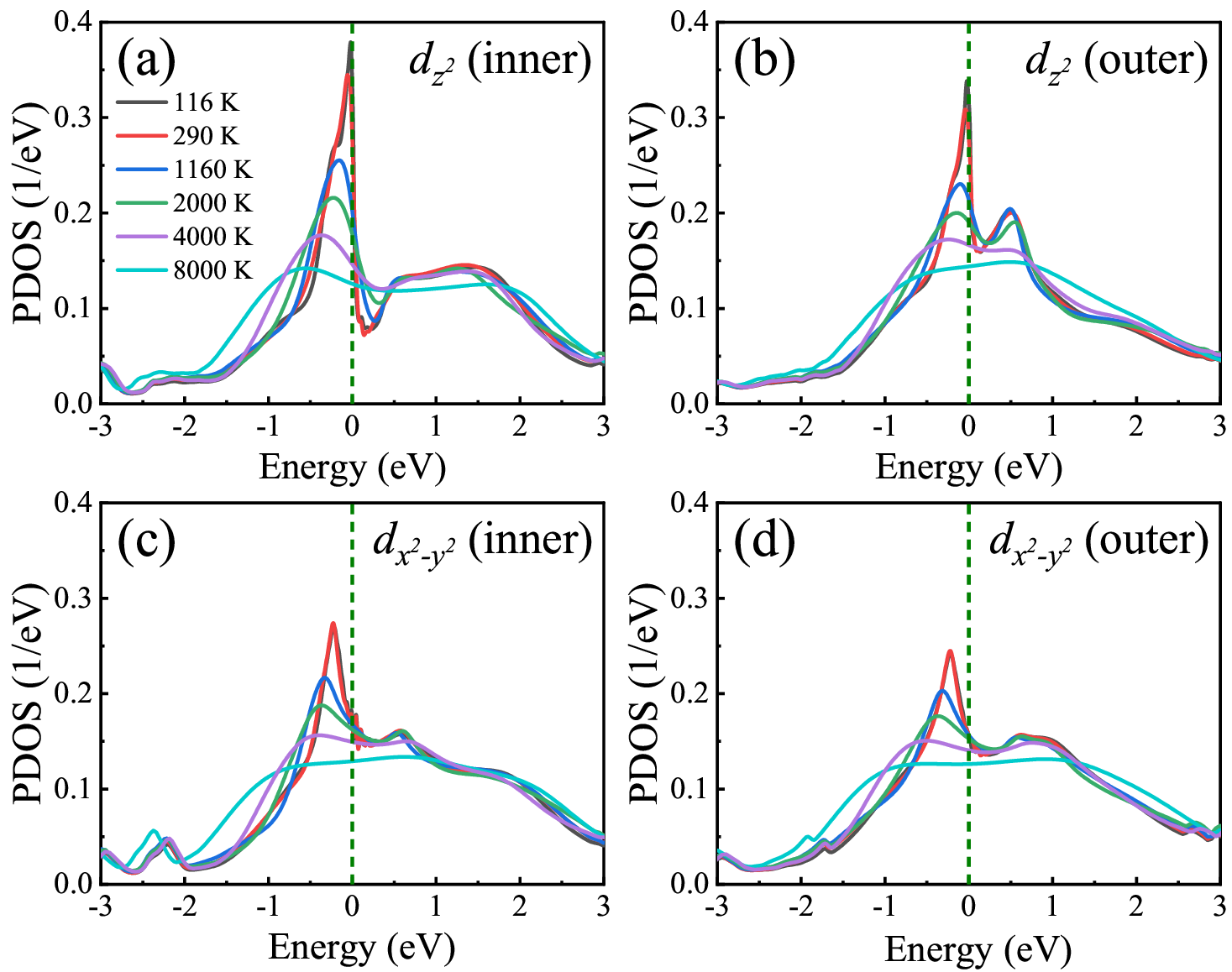}
  \caption{Projected density of states (PDOS), $A(\omega)$ $=$ $-1/\pi$${\rm Im}G(\omega)$, with several different temperatures for the inner-layer ${d_{z^2}}$ (a), outer-layer ${d_{z^2}}$ (b), inner-layer ${d_{x^2-y^2}}$ (c), and outer-layer ${d_{x^2-y^2}}$ orbitals (d), respectively.}
  \label{fig:s1}
\end{figure}

We studied the correlated real-frequency projected density of states (PDOS) at different temperatures, $A(\omega)=-1/\pi {\rm Im}G(\omega)$, for the inner- and outer-layer $e_g$ orbitals as shown in Fig.~\ref{fig:s1}. The correlated high temperature local spectra are characterized by a single broad hump and without pseudogap for both the inner- and outer-layer ${e_g}$ orbitals, suggesting that La4310 is not a Mott system. As temperature decreases, the magnitudes of the coherence peaks increase gradually and their positions move towards the Fermi level, implying that electrons become more coherent. The sharp peak near the Feimi level in the PDOS for the $d_{z^2}$ orbitals is much higher than that for the $d_{x^2-y^2}$ orbitals, which is because the bonding state energy band formed by the $d_{z^2}$ orbitals is very flat due to correlation induced band renormalization.

\section{Local susceptibilities}\label{sec:susc}
\begin{figure}[htb]
  \centering
  \includegraphics[width=6.0cm]{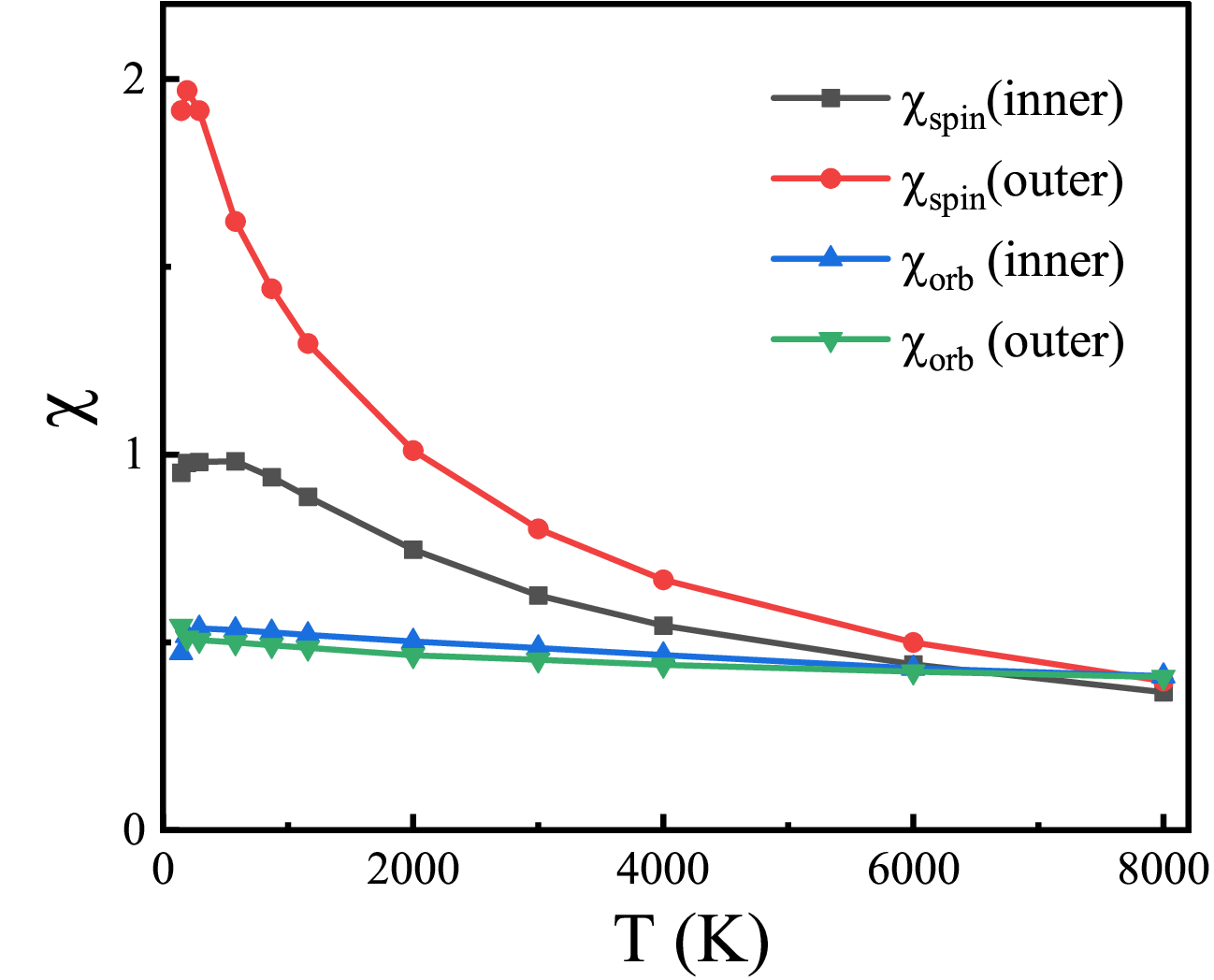}
  \caption{Static local susceptibilities of spin $\chi_{\rm spin}$ and orbital $\chi_{\rm orb}$ as functions of temperature.}
  \label{fig:s2}
\end{figure}

Next, we studied the temperature dependence of static local susceptibilities for spin, $\chi_{\rm spin}$ $=$ $\int_{0}^{\beta}$$\langle{S_z(\tau)S_z(0)} \rangle$$d\tau$, and orbital, $\chi_{\rm orb}$ $=$ $\int_{0}^{\beta}$$\langle O(\tau)O(0)\rangle$$d\tau$$-\beta{{\langle O \rangle}^2}$, as shown in Fig.~\ref{fig:s2}. The spin susceptibilities of the inner- and outer-layer Ni $e_g$ orbitals increase as the temperature decreases and saturate at approximately 290 and 190 K, respectively. For the inner-layer, the spin susceptibility appears to be constant below 290 K, consistent with Fermi liquid characteristics. In contrast, the outer-layer Ni atoms are difficult to be screened and have a high magnetic susceptibility. The local orbital susceptibilities are temperature independent. The screening of the orbital degrees of freedom has completed at a much higher temperature than that of the spin degrees of freedom, exhibiting spin-orbital separation \cite{deng2019nature}. This is a typical feature of the Hund correlation. 


\bibliography{lno4310hund}